\documentclass[letter]{spie}
\usepackage{amsmath,amsfonts,amssymb}
\usepackage{graphicx}
\usepackage[colorlinks=true, allcolors=blue]{hyperref}
\usepackage[square, numbers]{natbib}
\usepackage{enumitem}
\usepackage{multirow}
\usepackage{microtype}
\usepackage{mathptmx} 
\title{Colorectal Cancer Segmentation with Adaptive Augmentation and Multi-Resolution Ensemble Models}
\author{Ümit Mert Çağlar}
\author{Alptekin Temizel}
\affil{Graduate School of Informatics, METU, Ankara, Turkey}

\authorinfo{Further author information: (Send correspondence to Ümit Mert Çağlar)\\Ümit Mert Çağlar: E-mail: mecaglar@metu.edu.tr\\Alptekin Temizel: E-mail: atemizel@metu.edu.tr}

\begin{document} 
\maketitle

\begin{abstract}
Colorectal cancer (CRC) is the second most deadly and third most common cancer, and the leading cause of death among gastrointestinal cancers. Early diagnosis is crucial for the treatment of this cancer and increasing the survival rates.  Although CRC is more common in developed regions, its occurrence is also increasing in developing regions as well. CRC diagnosis relies on histopathology assessment post-biopsy. Automated deep learning algorithms can significantly reduce diagnosis time, enhancing efficiency and supporting timely clinical decisions.

We present an automated segmentation pipeline for whole-slide histopathology images that labels tumor grades 1–3 and normal mucosa. It utilizes dense prediction transformers with various encoder backbones, overlapping patches, and test-time augmentation. An adaptive augmentation policy, guided by large language models, further improves training. Top models were ensembled via soft voting, and mask refining post-processing steps, Gaussian blurring, morphological closing, and connected components analysis. On a colorectal cancer grade dataset, our method improved the F1 score from 62.92 to 69.84. Code is available here: \href{https://github.com/caglarmert/ICIP2025}{https://github.com/caglarmert/ICIP2025}
\end{abstract}


\section{Introduction}
\label{sec:intro}

Colorectal cancer (CRC) is a malignant tumor that develops in the tissues of the colon or rectum, frequently originating from abnormal growths known as polyps \cite{testa2018colorectal}. 

CRC develops gradually over the years through genetic mutations in precancerous polyps, influenced by environmental factors like diet, lifestyle, chronic inflammation (such as ulcerative colitis), and hereditary syndromes \cite{hossain2022colorectal}. Both inherited traits and lifestyle contribute to CRC risk. While common in developed countries \cite{sawicki2021review}, incidence and mortality rates are also rising in low and middle income countries due to lifestyle and economic transitions \cite{arnold2017global}.

CRC is the leading cause of death among gastrointestinal cancers, the second most deadly, and the third most common cancer worldwide \cite{granados2017colorectal}. Early detection through timely diagnosis and polyp removal is crucial for reducing incidence and mortality \cite{stintzing2014management, hossain2022colorectal}. Accurate tumor grading from biopsies informs prognosis and treatment, as CRC subtypes significantly impact patient outcomes \cite{guinney2015consensus}. However, a global shortage of specialists especially in low and middle income regions limits screening and early detection efforts \cite{jacobson2024colorectal, khan2023challenges}. Enhancing prevention, early diagnosis, and access to care is essential to address this issue \cite{arnold2017global}. Advances in artificial intelligence, particularly deep learning, offer significant potential to reduce CRC mortality through improved screening and early detection \cite{kanth2021screening}. With ongoing advancements, especially in transformer models, automated histopathological image segmentation can further accelerate CRC diagnosis \cite{liu2024review}.

In this work, we propose an automated segmentation pipeline for whole-slide pathology images. The main contributions can be summarized as follows:

\begin{itemize}[itemsep=0pt, topsep=0pt] 
    \item Comprehensive experiments to identify optimal architectures and models for CRC segmentation.
    \item Ablation study on loss functions, architectures, resolution, and adaptive augmentation.
    \item Ensemble strategy combining top-performing models to enhance CRC segmentation.
    \item Analysis of post-processing techniques and their effects on performance.
\end{itemize}

\section{Related Work}

Deep learning has seen successful applications in classification and segmentation of histopathology data, enhancing automation and accuracy in tissue analysis \cite{xu2017large, bera2019artificial}. Convolutional neural networks (CNNs) \cite{panic2020convolutional} and residual CNNs \cite{akilandeswari2022automatic} have shown promise in automating histopathological assessment. Recent advances in deep learning, particularly attention-based models, enable successful applications in histopathology image segmentation \cite{atabansi2023survey}. Many deep learning approaches, especially transformer models, are applied in medical imaging through different approaches. These approaches include single or multi scale, transformer, CNN or hybrid models \cite{shamshad2023transformers}. Encoder-decoder networks are predominant in image segmentation task thanks to their architecture extracting semantic features \cite{salpea2022medical}. Ensemble methods combine strengths of different models, often improving the performance as a result \cite{karthik2024ensemble} while image pre and post processing also improve the deep learning applications in digital pathology \cite{salvi2021impact}. Some promising image processing approaches are thresholding, Gaussian filtering, connected component analysis and morphological operations.

\section{Methodology}
\begin{figure}[ht]
    \centering
    \includegraphics[width=0.8\linewidth]{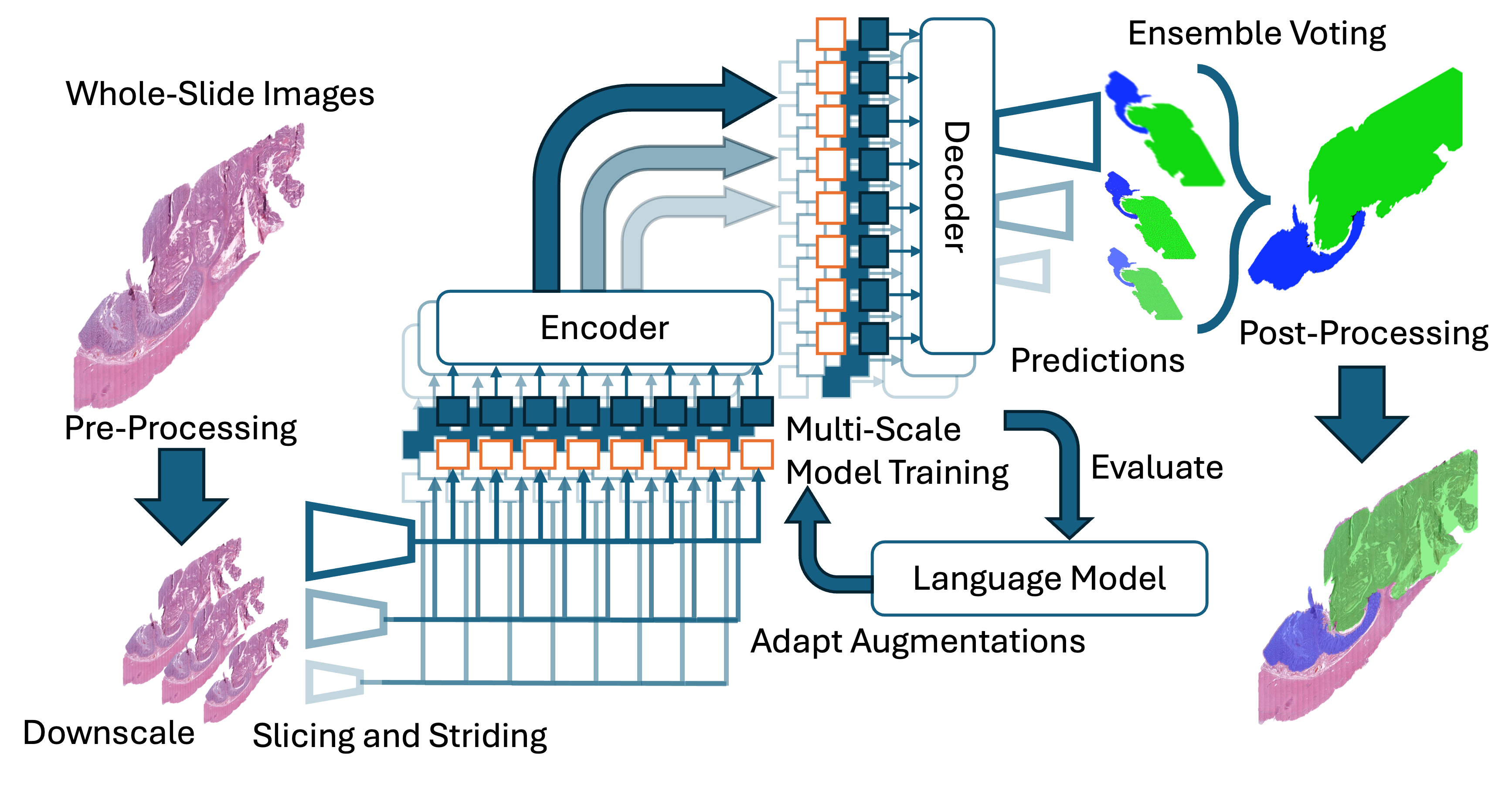}
    \caption{Overall flowchart of our methodology.}
    \label{fig:flowchart}
\end{figure}

Our method begins with downscaling whole-slide images into manageable smaller slices with varying overlap. Whole-slide imaging is a method of utilizing very high resolution images, such as histopathology imaging, that requires extra pre-processing steps \cite{kumar2020whole}. We train multiple encoder-decoder architectures for different resolution levels. We use language model guided adaptive augmentation policy optimization during training\cite{duru2024adaptive}. We ensemble best models according to the evaluation results and combine the prediction of these models with a soft-voting. We apply post-processing methods to finalize segmentation masks. This process is detailed in the Figure \ref{fig:flowchart}.

\subsection{Pre-processing}

To enable feasible training and experimentation, we employed a multi-resolution approach with three downscaling levels of $60\times$, $40\times$, and $20\times$. To enable memory-efficient processing, all WSIs were divided into non-overlapping $2048\times2048$ pixel patches, which were then individually downscaled in parallel. The downscaled patches were combined and saved in a lossless format to preserve image and label integrity, avoiding compression artifacts. We used nearest neighbor interpolation when downscaling masks to preserve class labels and bilinear interpolation for histopathology images. For our outputs, we employed area interpolation for per-pixel probability maps, as it effectively maintains regional structures \cite{roszkowiak2017survey}.
\newpage
\subsection{Model Architecture and Training}
We evaluated various models experimenting with different architectures, input scalings, patch sizes, strides, encoders, and loss functions to ensure robust generalization to test data. The architectures included Fully Connected Networks \cite{long2015fully}, U-Nets \cite{ronneberger2015u}, Segmentation Transformers \cite{xie2021segformer}, and Dense Prediction Transformers \cite{ranftl2021vision}. Encoder backbones comprised various sizes of Multi-Axis Vision Transformers \cite{tu2022maxvit}, EfficientNets \cite{tan2019efficientnet}, and Residual Networks \cite{he2016deep}. We trained models using multiple loss functions, including cross-entropy, Focal \cite{lin2017focal}, Lovasz \cite{berman2018lovasz}, Jaccard \cite{rahman2016optimizing}, Tversky \cite{salehi2017tversky}, and Dice \cite{milletari2016v}. All models were optimized with Adam with decoupled weight decay \cite{loshchilov2017decoupled} for stable and robust training. We used adaptive augmentation policy optimization with large language model (LLM) feedback \cite{duru2024adaptive} while training our models. This enabled us to adapt various models and configurations to the same dataset and task.

\subsection{Model Ensembling}

After training various segmentation models, we combined their outputs using ensemble strategies. A hard-voting ensemble was performed by selecting the majority class for each pixel among the top-performing models. Additionally, we extracted per-pixel class probability matrices and implemented a soft voting approach by aggregating the class-wise probabilities: summing the $top-N$ prediction probabilities across the best $M$ models and applying a class-specific bias vector that emphasizes tumor-related classes while minimizing background influence. The final segmentation masks were generated by taking the $argmax$ over these aggregated probability tensors.

\subsection{Post-Processing}

To refine the segmentation masks and reduce prediction noise, we implemented a three-stage post-processing pipeline. First, a Gaussian filter was applied to the probability maps to spatially smooth the predictions, which facilitated incorporating contextual information and reducing sharp, isolated errors at tumor boundaries. Next, morphological closing (dilation followed by erosion) was used to fill small holes and close narrow gaps, enhancing the shape integrity of tumor regions in accordance with histopathological patterns. Finally, connected components filtering removed small spurious regions by discarding connected areas below a predefined size threshold, addressing edge uncertainty and improving overall mask quality.

\section{Experimental Evaluation}

We utilized the Colorectal Cancer Tumor Grade Segmentation (CCTGS) dataset, comprising 103 whole-slide images with pixel-wise masks for three tumor grades and normal tissue \cite{arslan2025colorectal}. Among baseline models, the Swin Transformer \cite{liu2021swin} achieved the best performance of 62.92 F1 score. Our top 5 models achieved mean segmentation F1 scores ranging from 66.21 to 67.39. Using a straightforward majority hard-voting ensemble of these models increased the F1 score to 69.07. Further improvement was achieved with a top-3 soft-biased voting ensemble that combined prediction probability matrices. This approach leveraged the strengths of multiple models while reducing inconsistencies and class imbalance issues common in hard voting, ultimately raising the F1 score to 69.73. The post-processing techniques, Gaussian blurring, morphological closing, and connected components analysis, eliminated minor artifacts of our segmentation masks. This final refinement further enhanced the mean F1 score to 69.84.

\begin{table}[ht]
\centering
\caption{Comparison of baseline \cite{arslan2025colorectal} and our best-performing models.}
\begin{tabular}{lcccc|cccc}
& \multicolumn{4}{c|}{\textbf{Baseline} } & \multicolumn{4}{c}{\textbf{Ours}} \\
\textbf{Model} & \textbf{F1} & \textbf{Precision} & \textbf{Recall} & & \textbf{Model} & \textbf{F1} & \textbf{Precision} & \textbf{Recall} \\
\hline
SegFormer   & 45.07 & 56.44 & 50.52 & & UNet       & 56.79 & 60.20 & 54.70 \\
DeepLabv3+  & 52.31 & 51.52 & 62.35 & & FCN        & 59.81 & 57.37 & 64.68 \\
UNet        & 54.23 & 50.59 & 64.88 & & SegFormer  & 65.60 & 63.91 & 68.76 \\
ConvNext    & 56.24 & 54.70 & 61.42 & & DPT        & 67.39 & 63.44 & \textbf{72.31} \\
Swin        & 62.92 & 60.94 & 69.63 & & Ensemble   & \textbf{69.73} & \textbf{68.63} & 71.14 \\

\end{tabular}
\label{tab:comparative_results}
\end{table}

\begin{figure}[ht]
    \centering
    \includegraphics[width=0.6\linewidth]{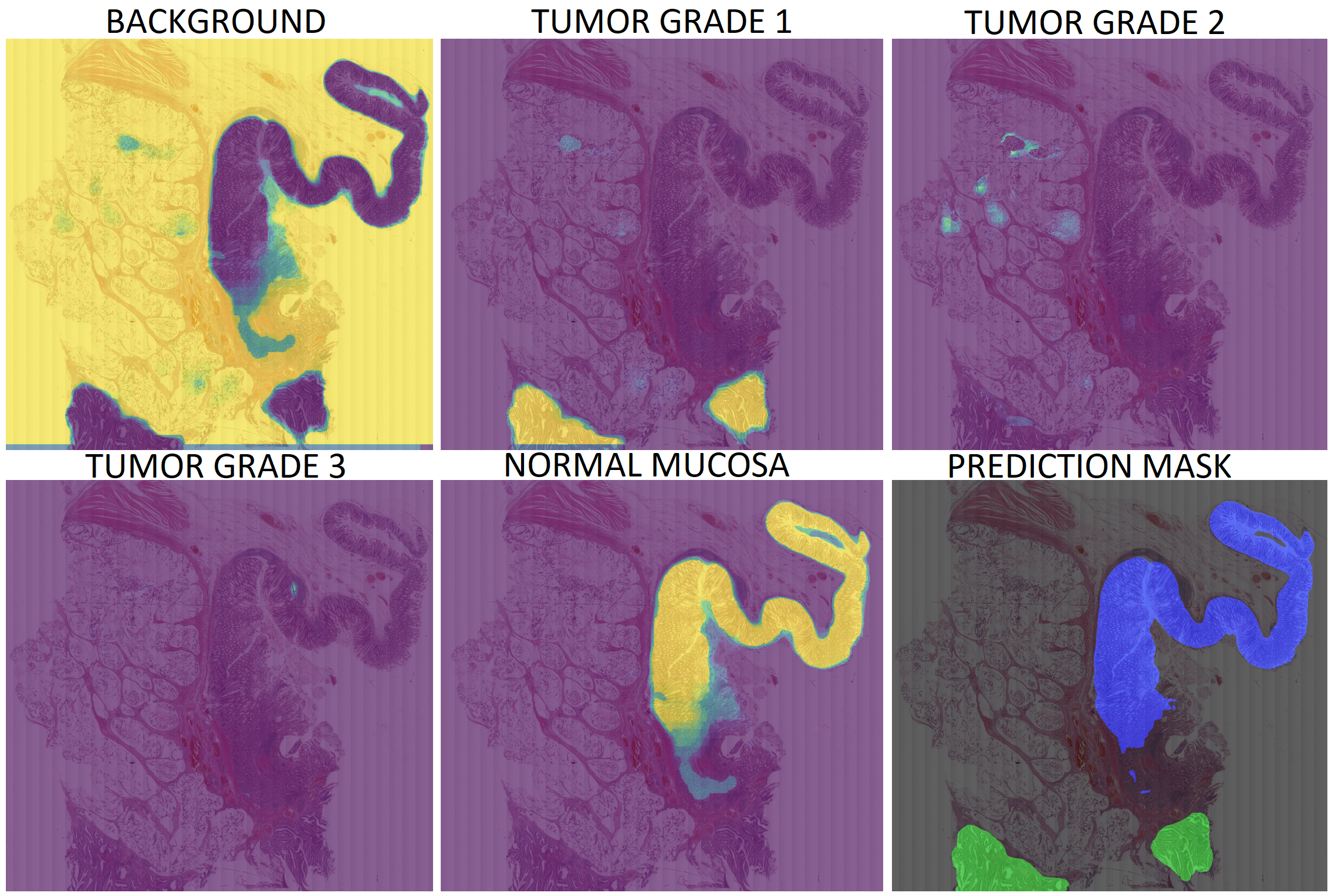}
    \caption{Overlays of the class probability heat-maps for five classes alongside the final segmentation masks.}
    \label{fig:overlays}
\end{figure}

Overlays on a sample histopathology image, showing five class activation maps alongside final segmentation masks, are displayed in Figure \ref{fig:overlays}. The overlays highlight that our model training converged to attend to different sections of the input images, effectively segmenting tumors from background.
\newpage
We've also used the EBHI-SEG dataset\cite{shi2023ebhi}, which contains 4,456 images across six colorectal tumor types. As shown in the Table \ref{tab:segmentation_results}, our approach consistently and substantially outperforms existing baselines, including U-Net\cite{ronneberger2015u}, Seg-Net\cite{badrinarayanan2017segnet}, and MedT\cite{valanarasu2021medical}, across nearly all classes and evaluation metrics. Notably, the models trained with adaptive augmentation achieve the highest scores, highlighting the robustness and effectiveness of our proposed method for colorectal tumor segmentation from histopathology images.

\begin{table}[ht]
\centering
\renewcommand{\arraystretch}{1.2}
\caption{Per-class segmentation performance comparison on EBHI-SEG dataset, across different models. Ours-1 is the best DPT model trained with static augmentation while Ours-2 is the best DPT model trained with adaptive augmentation.}
\begin{tabular}{cc}

\textbf{Normal} & \textbf{Polyp} \\
\begin{tabular}{lcccc}
Method & F1 Score & IoU & Precision & Recall \\
\hline 
U-Net   & 41.1 & 26.3 & 58.6 & 32.8 \\
Seg-Net & 77.7 & 68.4 & 89.5 & 75.8 \\
MedT    & 67.6 & 56.2 & 87.4 & 61.0 \\
Ours-1  & \textbf{93.8} & \textbf{88.3} & 94.0 & \textbf{93.6} \\
Ours-2  & \textbf{93.8} & \textbf{88.3} & \textbf{94.5} & 93.1 \\
\end{tabular} &
\begin{tabular}{lcccc}
Method & F1 Score & IoU & Precision & Recall \\
\hline 
U-Net   & 96.5 & \textbf{93.6} & 91.6 & 94.7 \\
Seg-Net & 93.7 & 88.6 & 91.0 & 96.5 \\
MedT    & 77.1 & 64.3 & 68.7 & 92.0 \\
Ours-1  & 96.2 & 92.6 & 95.9 & 96.4 \\ 
Ours-2  & \textbf{96.6} & 93.4 & \textbf{96.5} & \textbf{96.6} \\
\end{tabular} \\
\hline\hline
\textbf{High-grade IN} & \textbf{Low-grade IN} \\
\begin{tabular}{lcccc}
Method & F1 Score & IoU & Precision & Recall \\
\hline 
U-Net   & 89.5 & 81.6 & 84.7 & \textbf{96.1} \\
Seg-Net & 89.4 & 81.2 & 88.1 & 91.3 \\
MedT    & 82.4 & 70.7 & 74.0 & 93.8 \\
Ours-1  & 93.1 & 87.1 & 91.0 & 95.3 \\
Ours-2  & \textbf{94.1} & \textbf{88.8} & \textbf{93.9} & 94.3 \\
\end{tabular} &
\begin{tabular}{lcccc}
Method & F1 Score & IoU & Precision & Recall \\
\hline 
U-Net   & 91.1 & 84.9 & 87.9 & 95.3 \\
Seg-Net & 92.4 & 86.4 & 88.0 & \textbf{97.7} \\
MedT    & 88.9 & 80.8 & 87.6 & 91.6 \\
Ours-1  & 96.3 & 92.8 & 95.8 & 96.7 \\ 
Ours-2  & \textbf{96.4} & \textbf{93.1} & \textbf{95.9} & 96.9 \\
\end{tabular} \\
\hline\hline
\textbf{Adenocarcinoma} & \textbf{Serrated adenoma} \\
\begin{tabular}{lcccc}
Method & F1 Score & IoU & Precision & Recall \\
\hline 
U-Net   & 88.7 & 80.8 & 85.0 & 95.0 \\
Seg-Net & 86.5 & 77.5 & 79.2 & \textbf{97.7} \\
MedT    & 73.5 & 59.5 & 68.2 & 86.4 \\
Ours-1  & 90.6 & 82.9 & \textbf{91.4} & 89.9 \\
Ours-2  & \textbf{91.1} & \textbf{83.7} & 90.7 & 91.5 \\
\end{tabular} &
\begin{tabular}{lcccc}
Method & F1 Score & IoU & Precision & Recall \\
\hline 
U-Net   & 93.8 & 88.6 & 89.9 & \textbf{98.3} \\
Seg-Net & 90.7 & 83.2 & 85.9 & 96.3 \\
MedT    & 67.0 & 50.9 & 89.6 & 54.4 \\
Ours-1  & 93.8 & 88.2 & 94.9 & 92.6 \\ 
Ours-2  & \textbf{95.5} & \textbf{91.4} & \textbf{96.7} & 94.3 \\ 
\end{tabular} \\

\end{tabular}
\label{tab:segmentation_results}
\end{table}

\clearpage
\newpage
\section{Ablation Study}

To optimize histopathology image segmentation, we trained 245 models over 1,500+ hours, testing six loss functions (CE, Jaccard, Tversky, Focal, Lovasz, Dice), three optimizers (Adam, AdamW, SGD), four architectures with ten encoder backbones, and three input resolutions.

\textbf{Loss Functions:} Performance evaluation of models trained with different loss functions are summarized in Table \ref{tab:combined_comparison}. These results stem from extensive hyper-parameter searches, representing the best models trained with each loss. The cross-entropy (CE) loss performed the worst, likely due to the class imbalance caused by the dominance of background pixels; CE does not account for this imbalance, leading to inferior performance. The most effective loss was Dice, which directly optimizes the F1 score, making it well-suited for our goal. Tversky loss ranked second, with a slight bias towards precision (0.6) over recall (0.4), resulting in better precision at the expense of recall compared to Dice. Jaccard loss was a close third, achieving slightly lower F1 scores but marginally higher IoU than Tversky. CE and Focal losses perform poorly due to their emphasis on cross-entropy minimization. Lovasz loss is similarly unsuitable, showing only slight improvement over Focal. Conversely, Jaccard, Tversky, and Dice losses provide the most consistent and robust results.

\textbf{Architecture Search:} We evaluated several architectures, including Fully Connected Networks (FCN) \cite{long2015fully}, U-Net \cite{ronneberger2015u}, Segmentation Transformers (SegFormer) \cite{xie2021segformer}, and Dense Prediction Transformers (DPT) \cite{ranftl2021vision}, with various backbones. Among these, SegFormer and DPT architectures demonstrated superior performance, surpassing the literature baseline (Table \ref{tab:combined_comparison}).

\textbf{Downscaling Levels:} We downscaled the original whole-slide images by various factors. The $20\times$ downscaling yields images with a resolution four times higher than those downscaled by $40\times$, and nine times higher than those downscaled by $60\times$, leading to longer training times. As a trade-off, the $40\times$ downscaling serves as an intermediate level that balances detail and computational efficiency.

We trained models at each resolution level and observed comparable performance across all levels. This underscores the importance of careful architecture selection and hyperparameter optimization during training.

The best models for each resolution are summarized in Table \ref{tab:combined_comparison}. All were based on the DPT architecture with MaxViT encoder backbones. Specifically, the $60\times$ downscale model was trained using Dice loss with a MaxViT-Large encoder and a $224\times224$ window size. The $20\times$ downscale model also used Dice loss with MaxViT-Large and a $512\times512$ window size, while the $40\times$ downscale model employed Jaccard loss with the same encoder and window size. These results are consistent with our earlier findings regarding the effectiveness of different loss functions.

\textbf{Adaptive Augmentation:} We implemented an adaptive augmentation policy optimization approach that employs a large language model (LLM) as the core of a feedback loop. This feedback loop considers the current augmentation strategy, evaluation metrics, and training details such as the model architecture and hyperparameters to generate updates to the augmentation policy. We trained models both with and without this adaptive augmentation strategy to evaluate its effectiveness, as evident from the Table \ref{tab:combined_comparison}.

\begin{table}[ht]
\centering
\caption{Performance comparison across loss functions, architectures, resolutions, and augmentation strategies.}
\begin{tabular}{llcccc}
\textbf{Category} & \textbf{Configuration} & \textbf{F1 Score} & \textbf{IoU} & \textbf{Precision} & \textbf{Recall} \\
\hline
\multirow{6}{*}{Loss Function} 
& CE         & 61.69 & 47.91 & 60.31 & 63.79 \\
& Focal      & 65.53 & 51.44 & 61.42 & 72.19 \\
& Lovasz     & 65.91 & 51.77 & 64.95 & 70.49 \\
& Jaccard    & 66.82 & 52.53 & 63.60 & 71.06 \\
& Tversky    & 66.87 & 52.47 & \textbf{69.28} & 67.51 \\
& Dice       & \textbf{67.39} & \textbf{53.15} & 63.44 & \textbf{72.31} \\
\hline
\multirow{5}{*}{Architecture} 
& UNet-ResNet101& 56.79 & 43.32 & 60.20 & 54.70 \\
& FCN-ResNet101& 59.81 & 45.89 & 57.37 & 64.68 \\
& Segformer-efficientnet-b7& 65.60 & 52.04 & 63.91 & 68.76 \\
& DPT-MaxVit-224& 67.11 & \textbf{53.42} & \textbf{65.78} & 70.46 \\
& DPT-MaxVit-512& \textbf{67.39} & 53.15 & 63.44 & \textbf{72.31} \\
\hline
\multirow{3}{*}{Resolution} 
& Downscale $60\times$ & 67.11 & \textbf{53.42} & \textbf{65.78} & 70.46 \\
& Downscale $40\times$ & 66.82 & 52.53 & 63.60 & 71.06 \\
& Downscale $20\times$ & \textbf{67.39} & 53.15 & 63.44 & \textbf{72.31} \\
\hline
\multirow{4}{*}{Augmentation Strategy} 
& Static (Model 1)   & 57.76 & 43.89 & 53.30 & \textbf{65.61} \\
& Adaptive (Model 1) & \textbf{59.48} & \textbf{45.81} & \textbf{56.55} & 63.28 \\
& Static (Model 2)   & 63.08 & 48.83 & 62.10 & 65.44 \\
& Adaptive (Model 2) & \textbf{67.39} & \textbf{53.15} & \textbf{63.44} & \textbf{72.31} \\
\hline
\multirow{4}{*}{Ensemble Strategy} 
&Best Single Scale Model& 67.39 & 53.15 & 63.44 & 72.31 \\
&Hard-voting Ensemble & 69.07 & 55.00 & 65.41 & \textbf{73.89}\\
&Soft-voting Ensemble* & 69.73 & 55.82 & \textbf{68.63} & 71.14 \\
&* with post-processing & \textbf{69.84} & \textbf{55.83} & 68.62 & 71.66\\
\end{tabular}
\label{tab:combined_comparison}
\end{table}

Extensive ablation studies, detailed in the Table \ref{tab:combined_comparison} reveal that:
\begin{itemize}[itemsep=0pt, topsep=0pt] 
 \item Loss functions such as Dice, Tversky, and Jaccard outperform standard cross entropy and focal loss. 
 \item Transformer-based segmentation architectures with pre-trained transformer encoders yield superior performance compared to traditional U-Net or SegFormer architectures with CNN encoders. 
 \item The LLM-guided adaptive augmentation strategy improves model performance. 
 \item Ensemble strategies boost segmentation performance significantly, and soft-voting is better than hard-voting.
 \item Post-processing techniques marginally improve results without additional training costs.
\end{itemize}
\newpage
\section{Conclusions}
We developed an image segmentation pipeline for multi-class tumor grading in colorectal cancer histopathology images. Our approach, based on transformer models with multi-scale ensemble methods and adaptive augmentation policy optimization, achieved state-of-the-art performance, surpassing F1 scores of existing baselines by over 6 points. Specifically, we employed dense prediction transformers (DPT) with MaxViT encoder backbones and extensive patch overlapping to maximize training data.

To improve results, we applied test-time augmentation by generating predictions under different conditions and ensembled them with multiple models trained at various resolutions. Using different resolutions enabled models to capture both fine and coarse details, resulting in diverse segmentation probability matrices. These were combined via soft voting, biased toward tumor classes, to create robust masks. Finally, post-processing techniques further refined the segmentation results.

Future work includes enhancing ensemble methods with binary segmentation models in a one-vs-all setting. Test-time augmentation can be expanded with additional strategies or integrated into a separate adaptive feedback loop. While we have used $20$, $40$ and $60\times$ downscaled images, future studies with greater resources could leverage full-resolution images and larger models to capture details lost during downsampling.

\section*{Acknowledgment}

This work has been supported by Middle East Technical University Scientific Research Projects Coordination Unit under grant number ADEP-704-2024-11486 and TUBITAK 2224-A-Grant Program. The experiments reported in this work were fully performed at TUBITAK ULAKBIM, High Performance and Grid Computing Center (TRUBA) and at The Artificial Intelligence and Big Data Analytics Laboratory at METU-DTX.
\newpage
\bibliography{report}

@article{roszkowiak2017survey,
  title={Survey: interpolation methods for whole slide image processing},
  author={Roszkowiak, L and Korzynska, A and Zak, J and Pijanowska, Dorota and Swiderska-Chadaj, Z and Markiewicz, Tomasz},
  journal={Journal of microscopy},
  volume={265},
  number={2},
  pages={148--158},
  year={2017},
  publisher={Wiley Online Library}
}

@article{badrinarayanan2017segnet,
  title={Segnet: A deep convolutional encoder-decoder architecture for image segmentation},
  author={Badrinarayanan, Vijay and Kendall, Alex and Cipolla, Roberto},
  journal={IEEE transactions on pattern analysis and machine intelligence},
  volume={39},
  number={12},
  pages={2481--2495},
  year={2017},
  publisher={IEEE}
}

@inproceedings{valanarasu2021medical,
  title={Medical transformer: Gated axial-attention for medical image segmentation},
  author={Valanarasu, Jeya Maria Jose and Oza, Poojan and Hacihaliloglu, Ilker and Patel, Vishal M},
  booktitle={Medical image computing and computer assisted intervention--MICCAI 2021: 24th international conference, Strasbourg, France, September 27--October 1, 2021, proceedings, part I 24},
  pages={36--46},
  year={2021},
  organization={Springer}
}

@article{stintzing2014management,
  title={Management of colorectal cancer},
  author={Stintzing, Sebastian},
  journal={F1000prime reports},
  volume={6},
  pages={108},
  year={2014}
}

@article{granados2017colorectal,
  title={Colorectal cancer: a review},
  author={Granados-Romero, Juan Jos{\'e} and Valderrama-Trevi{\~n}o, Alan Isaac and Contreras-Flores, Ericka Hazzel and Barrera-Mera, Baltazar and Herrera Enr{\'\i}quez, Miguel and Uriarte-Ru{\'\i}z, Karen and Ceballos-Villalba, Jes{\'u}s Carlos and Estrada-Mata, Aranza Guadalupe and Alvarado Rodr{\'\i}guez, Cristopher and Arauz-Pe{\~n}a, Gerardo},
  journal={Int J Res Med Sci},
  volume={5},
  number={11},
  pages={4667},
  year={2017}
}

@article{akilandeswari2022automatic,
  title={Automatic detection and segmentation of colorectal cancer with deep residual convolutional neural network},
  author={Akilandeswari, A and Sungeetha, D and Joseph, Christeena and Thaiyalnayaki, K and Baskaran, K and Jothi Ramalingam, R and Al-Lohedan, Hamad and Al-Dhayan, Dhaifallah M and Karnan, Muthusamy and Meansbo Hadish, Kibrom},
  journal={Evidence-Based Complementary and Alternative Medicine},
  volume={2022},
  number={1},
  pages={3415603},
  year={2022},
  publisher={Wiley Online Library}
}

@inproceedings{panic2020convolutional,
  title={A Convolutional Neural Network based system for Colorectal cancer segmentation on MRI images},
  author={Panic, Jovana and Defeudis, Arianna and Mazzetti, Simone and Rosati, Samanta and Giannetto, Giuliana and Vassallo, Lorenzo and Regge, Daniele and Balestra, Gabriella and Giannini, Valentina},
  booktitle={Int. Conf. IEEE Engineering in Medicine \& Biology Society (EMBC)},
  year={2020},
}

@article{bera2019artificial,
  title={Artificial intelligence in digital pathology—new tools for diagnosis and precision oncology},
  author={Bera, Kaustav and Schalper, Kurt A and Rimm, David L and Velcheti, Vamsidhar and Madabhushi, Anant},
  journal={Nature reviews Clinical oncology},
  volume={16},
  number={11},
  pages={703--715},
  year={2019},
  publisher={Nature Publishing Group UK London}
}

@article{guinney2015consensus,
  title={The consensus molecular subtypes of colorectal cancer},
  author={Guinney, Justin and Dienstmann, Rodrigo and Wang, Xin and De Reynies, Aur{\'e}lien and Schlicker, Andreas and Soneson, Charlotte and Marisa, Laetitia and Roepman, Paul and Nyamundanda, Gift and Angelino, Paolo and others},
  journal={Nature medicine},
  volume={21},
  number={11},
  pages={1350--1356},
  year={2015},
  publisher={Nature Publishing Group US New York}
}

@article{shi2023ebhi,
  title={EBHI-Seg: A novel enteroscope biopsy histopathological hematoxylin and eosin image dataset for image segmentation tasks},
  author={Shi, Liyu and Li, Xiaoyan and Hu, Weiming and Chen, Haoyuan and Chen, Jing and Fan, Zizhen and Gao, Minghe and Jing, Yujie and Lu, Guotao and Ma, Deguo and others},
  journal={Frontiers in Medicine},
  volume={10},
  pages={1114673},
  year={2023},
  publisher={Frontiers Media SA}
}

@article{arslan2025colorectal,
  title={Colorectal Cancer Tumor Grade Segmentation: A new dataset and baseline results},
  author={Arslan, Duygu and Sehlaver, Sina and Guder, Erce and Temena, Mehmet Arda and Bahcekapili, Alper and Ozdemir, Umut and Turkay, Duriye Ozer and Guner, Gunes and Guresci, Servet and Sokmensuer, Cenk and others},
  journal={Heliyon},
  volume={11},
  year={2025},
  publisher={Elsevier}
}

@article{loshchilov2017decoupled,
  title={Decoupled weight decay regularization},
  author={Loshchilov, Ilya and Hutter, Frank},
  journal={arXiv preprint arXiv:1711.05101},
  year={2017}
}

@inproceedings{he2016deep,
  title={Deep residual learning for image recognition},
  author={He, Kaiming and Zhang, Xiangyu and Ren, Shaoqing and Sun, Jian},
  booktitle={CVPR},
  year={2016}
}

@inproceedings{tan2019efficientnet,
  title={Efficientnet: Rethinking model scaling for convolutional neural networks},
  author={Tan, Mingxing and Le, Quoc},
  booktitle={ICML},
  pages={6105--6114},
  year={2019},
  organization={PMLR}
}

@article{kumar2020whole,
  title={Whole slide imaging (WSI) in pathology: current perspectives and future directions},
  author={Kumar, Neeta and Gupta, Ruchika and Gupta, Sanjay},
  journal={Journal of digital imaging},
  volume={33},
  number={4},
  pages={1034--1040},
  year={2020},
  publisher={Springer}
}

@article{salvi2021impact,
  title={The impact of pre-and post-image processing techniques on deep learning frameworks: A comprehensive review for digital pathology image analysis},
  author={Salvi, Massimo and Acharya, U Rajendra and Molinari, Filippo and Meiburger, Kristen M},
  journal={Computers in Biology and Medicine},
  volume={128},
  pages={104129},
  year={2021},
  publisher={Elsevier}
}

@inproceedings{salpea2022medical,
  title={Medical image segmentation: A review of modern architectures},
  author={Salpea, Natalia and Tzouveli, Paraskevi and Kollias, Dimitrios},
  booktitle={European Conference on Computer Vision},
  pages={691--708},
  year={2022},
  organization={Springer}
}

@article{karthik2024ensemble,
  title={Ensemble-based multimodal medical imaging fusion for tumor segmentation},
  author={Karthik, A and Hamatta, Hatem SA and Patthi, Sridhar and Krubakaran, C and Pradhan, Abhaya Kumar and Rachapudi, Venubabu and Shuaib, Mohammed and Rajaram, A},
  journal={Biomedical Signal Processing and Control},
  volume={96},
  pages={106550},
  year={2024},
  publisher={Elsevier}
}

@article{shamshad2023transformers,
  title={Transformers in medical imaging: A survey},
  author={Shamshad, Fahad and Khan, Salman and Zamir, Syed Waqas and Khan, Muhammad Haris and Hayat, Munawar and Khan, Fahad Shahbaz and Fu, Huazhu},
  journal={Medical image analysis},
  volume={88},
  pages={102802},
  year={2023},
  publisher={Elsevier}
}

@article{atabansi2023survey,
  title={A survey of Transformer applications for histopathological image analysis: New developments and future directions},
  author={Atabansi, Chukwuemeka Clinton and Nie, Jing and Liu, Haijun and Song, Qianqian and Yan, Lingfeng and Zhou, Xichuan},
  journal={BioMedical Engineering OnLine},
  volume={22},
  number={1},
  pages={96},
  year={2023},
  publisher={Springer}
}

@inproceedings{liu2021swin,
  title={Swin transformer: Hierarchical vision transformer using shifted windows},
  author={Liu, Ze and Lin, Yutong and Cao, Yue and Hu, Han and Wei, Yixuan and Zhang, Zheng and Lin, Stephen and Guo, Baining},
  booktitle={ICCV},
  pages={10012--10022},
  year={2021}
}

@article{duru2024adaptive,
  title={Adaptive Augmentation Policy Optimization with {LLM} Feedback},
  author={Duru, Ant and Temizel, Alptekin},
  journal={arXiv preprint arXiv:2410.13453},
  year={2024}
}

@inproceedings{ranftl2021vision,
  title={Vision transformers for dense prediction},
  author={Ranftl, Ren{\'e} and Bochkovskiy, Alexey and Koltun, Vladlen},
  booktitle={ICCV},
  pages={12179--12188},
  year={2021}
}

@inproceedings{tu2022maxvit,
  title={Maxvit: Multi-axis vision transformer},
  author={Tu, Zhengzhong and Talebi, Hossein and Zhang, Han and Yang, Feng and Milanfar, Peyman and Bovik, Alan and Li, Yinxiao},
  booktitle={ECCV},
  pages={459--479},
  year={2022},
  organization={Springer}
}

@article{kanth2021screening,
  title={Screening and prevention of colorectal cancer},
  author={Kanth, Priyanka and Inadomi, John M},
  journal={Bmj},
  volume={374},
  year={2021},
  publisher={British Medical Journal Publishing Group}
}

@article{hossain2022colorectal,
  title={Colorectal cancer: a review of carcinogenesis, global epidemiology, current challenges, risk factors, preventive and treatment strategies},
  author={Hossain, Md Sanower and Karuniawati, Hidayah and Jairoun, Ammar Abdulrahman and Urbi, Zannat and Ooi, Der Jiun and John, Akbar and Lim, Ya Chee and Kibria, KM Kaderi and Mohiuddin, AKM and Ming, Long Chiau and others},
  journal={Cancers},
  volume={14},
  number={7},
  pages={1732},
  year={2022},
  publisher={MDPI}
}

@article{testa2018colorectal,
  title={Colorectal cancer: genetic abnormalities, tumor progression, tumor heterogeneity, clonal evolution and tumor-initiating cells},
  author={Testa, Ugo and Pelosi, Elvira and Castelli, Germana},
  journal={Medical Sciences},
  volume={6},
  number={2},
  pages={31},
  year={2018},
  publisher={MDPI}
}

@article{arnold2017global,
  title={Global patterns and trends in colorectal cancer incidence and mortality},
  author={Arnold, Melina and Sierra, M{\'o}nica S and Laversanne, Mathieu and Soerjomataram, Isabelle and Jemal, Ahmedin and Bray, Freddie},
  journal={Gut},
  volume={66},
  number={4},
  pages={683--691},
  year={2017},
  publisher={BMJ Publishing Group}
}

@inproceedings{liu2024review,
  title={A Review of Colorectal Cancer Histopathology Image Segmentation Using Deep Learning Methods},
  author={Liu, Tengxiao and Yang, Huiya and Huang, Jiayi and Zhou, Yihan},
  booktitle={Int. Conf. Control and Robotics (ICCR)},
  year={2024},
}

@article{sawicki2021review,
  title={A review of colorectal cancer in terms of epidemiology, risk factors, development, symptoms and diagnosis},
  author={Sawicki, Tomasz and Ruszkowska, Monika and Danielewicz, Anna and Nied{\'z}wiedzka, Ewa and Ar{\l}ukowicz, Tomasz and Przyby{\l}owicz, Katarzyna E},
  journal={Cancers},
  volume={13},
  number={9},
  pages={2025},
  year={2021},
  publisher={Mdpi}
}

@article{khan2023challenges,
  title={Challenges in the management of colorectal cancer in low-and middle-income countries},
  author={Khan, Shah Zeb and Lengyel, Csongor Gy{\"o}rgy},
  journal={Cancer Treatment and Research Communications},
  volume={35},
  pages={100705},
  year={2023},
  publisher={Elsevier}
}

@article{jacobson2024colorectal,
  title={Colorectal cancer outcomes: a comparative review of resource-limited settings in low-and middle-income countries and rural America},
  author={Jacobson, Clare E and Harbaugh, Calista M and Agbedinu, Kwabena and Kwakye, Gifty},
  journal={Cancers},
  volume={16},
  number={19},
  pages={3302},
  year={2024},
  publisher={MDPI}
}

@article{xu2017large,
  title={Large scale tissue histopathology image classification, segmentation, and visualization via deep convolutional activation features},
  author={Xu, Yan and Jia, Zhipeng and Wang, Liang-Bo and Ai, Yuqing and Zhang, Fang and Lai, Maode and Chang, Eric I-Chao},
  journal={BMC bioinformatics},
  volume={18},
  pages={1--17},
  year={2017},
  publisher={Springer}
}

@inproceedings{long2015fully,
  title={Fully convolutional networks for semantic segmentation},
  author={Long, Jonathan and Shelhamer, Evan and Darrell, Trevor},
  booktitle={CVPR},
  pages={3431--3440},
  year={2015}
}

@inproceedings{milletari2016v,
  title={V-net: Fully convolutional neural networks for volumetric medical image segmentation},
  author={Milletari, Fausto and Navab, Nassir and Ahmadi, Seyed-Ahmad},
  booktitle={Int. Conf. 3D vision (3DV)},
  pages={565--571},
  year={2016},
  organization={Ieee}
}

@inproceedings{salehi2017tversky,
  title={Tversky loss function for image segmentation using 3D fully convolutional deep networks},
  author={Salehi, Seyed Sadegh Mohseni and Erdogmus, Deniz and Gholipour, Ali},
  booktitle={International workshop on machine learning in medical imaging},
  pages={379--387},
  year={2017},
  organization={Springer}
}

@inproceedings{rahman2016optimizing,
  title={Optimizing intersection-over-union in deep neural networks for image segmentation},
  author={Rahman, Md Atiqur and Wang, Yang},
  booktitle={International symposium on visual computing},
  pages={234--244},
  year={2016},
  organization={Springer}
}

@inproceedings{berman2018lovasz,
  title={The lov{\'a}sz-softmax loss: A tractable surrogate for the optimization of the intersection-over-union measure in neural networks},
  author={Berman, Maxim and Triki, Amal Rannen and Blaschko, Matthew B},
  booktitle={CVPR},
  pages={4413--4421},
  year={2018}
}

@inproceedings{lin2017focal,
  title={Focal loss for dense object detection},
  author={Lin, Tsung-Yi and Goyal, Priya and Girshick, Ross and He, Kaiming and Doll{\'a}r, Piotr},
  booktitle={ICCV},
  pages={2980--2988},
  year={2017}
}

@article{xie2021segformer,
  title={SegFormer: Simple and efficient design for semantic segmentation with transformers},
  author={Xie, Enze and Wang, Wenhai and Yu, Zhiding and Anandkumar, Anima and Alvarez, Jose M and Luo, Ping},
  journal={NeurIPS},
  volume={34},
  pages={12077--12090},
  year={2021}
}

@inproceedings{ronneberger2015u,
  title={U-net: Convolutional networks for biomedical image segmentation},
  author={Ronneberger, Olaf and Fischer, Philipp and Brox, Thomas},
  booktitle={Medical image computing and computer-assisted intervention--MICCAI},
  year={2015},
  organization={Springer}
}
\bibliographystyle{spiebib}


\end{document}